# Dissension or consensus? Management and Business Research in Latin America and the Caribbean

Julián D. Cortés

*Universidad del Rosario, Colombia*

*Fudan University, China*

E-mail: julian.cortess@urosario.edu.co

**Abstract**

This study presents longitudinal evidence on the dissension of Management and Business Research (MBR) in Latin America and the Caribbean (LAC). It looks after intellectual bridges linking clusters among such dissension. It was implemented a coword network analysis to a sample of 12,000+ articles published by authors from LAC during 1998-2017. Structural network scores showed an increasing number of keywords and mean degree but decreasing modularity and density. The intellectual bridges were those of the cluster formed by disciplines/fields that tend toward consensus (e.g., *mathematical models*) and not by core MBR subjects (e.g., *strategic planning*).

**Introduction**

The Winchester Mystery House has been used as an analogy for Management and Business related Research (MBR) —a purposeless although expensive entity (Davis, 2015). The hypothesis of the Hierarchy of Sciences states that while some sciences/disciplines studying simple phenomena (i.e., cells' functioning) will tend toward consensus, others studying complex phenomena (i.e., human collective behavior) will tend toward dissension (Fanelli & Glänzel, 2013). Thus, the latter path should be shaping MBR. Bottom line, dissension fragments research findings to advance on —a significant downsize effect for regions with scarce R&D resources such as Latin America and the Caribbean (LAC) (Cortés-Sánchez, 2019).

This study aims two-fold: i) to explore such dissension in MBR in LAC; ii) and to identify potential *intellectual bridges* to pooling efforts and multiply impact. I implemented the coword analysis (Callon et al., 1983). Such a science mapping (SM) technique enables stakeholders to map individual/institutional/national research and knowledge competencies (Shiffrin & Börner, 2004). Findings would be of interest to researchers in MBR, business schools, and funders by presenting the first comprehensive regional study for MBR to identify research clusters, topics as intellectual bridges between disciplines, and their evolution since 1998.

**Methodology**

This study modeled four coword networks for the periods 1998-2002, 2003-2007, 2008-2012, and 2013-2017. Dissension paths were explored by computing and comparing macro, meso, and micro structural network scores. Dataset and high-resolution figures are available in open access for replication and further use (Baker & Penny, 2016): http://bit.ly/2QyDJNP

*Data*

Bibliographic data was sourced from Scopus due to its journal coverage and LAC authors' involvement (Mongeon & Paul-Hus, 2016). The query searched for articles published from 1998 to 2017 by at least one author affiliated with any LAC institution in SCImago's *business, management, and accounting* subject area (SCImago, 2020). The final sample consisted of 12,149 articles after removing a journal with predatory features (Cortés-Sánchez, 2019) and articles with missing data. Table 1 presents the bibliometric descriptives for the top-10 LAC countries by output.

**Table 1 Bibliometric descriptives for the top-10 LAC countries by output.**

| | | | | | | |
|---|---|---|---|---|---|---|
| Total articles | 12,149 | | | | | |
| Av. citation per article | 14.02 | | | | | |
| Authors | 20,694 | | | | | |
| Authors per document | 1.7 | | | | | |
| Annual growth % | 17.2 | | | | | |
| Corresponding author's countries | Articles | % Sample | MCP | Av. citation per article | Most relevant sources | % Sample |
| Brazil | 3,256 | 40% | 23 | 14.0 | *Información Tecnológica* | 7.8% |
| Mexico | 874 | 10% | 11 | 14.3 | *J. of Cleaner Production* | 5.6% |
| Colombia | 717 | 8% | 5 | 8.4 | *Gestão & Produção* | 4.1% |
| Chile | 644 | 8% | 6 | 17.0 | *R. de Administração de Empresas* | 2.4% |
| Argentina | 427 | 5.3% | 4 | 11.3 | *J. of Technology Management and Innovation* | 2.2% |
| Venezuela | 129 | 1.6% | 4 | 3.8 | *J. of Business Research* | 1.9% |
| Peru | 96 | 1.2% | 2 | 15.7 | *R. Venezolana de Gerencia* | 1.9% |
| Costa Rica | 63 | 0.7% | 0 | 16.4 | *Estudios Gerenciales* | 1.8% |
| Uruguay | 55 | 0.6% | 0 | 12.5 | *R. Brasileira de Gestão de Negocios* | 1.8% |
| Cuba | 43 | 0.5% | 0 | 12.9 | *International J. of Production Research* | 1.4% |

Source: elaborated by the author based on Scopus (2020). Note: MCP: Multi-Country Publication.

*Methods and software*

Coword analysis enables putting together the conceptual structure based on the co-occurrence of articles keywords (Callon et al., 1983). *KeyWords Plus* method generates key-terms based on the articles' title and the references cited appearing more than twice (Clarivate Analytics, n.d.). A link connects two keywords (i.e., nodes) if both appear in the same research article (i.e., edge). The scores computed for the coword networks were: i) macro: density, mean degree, modularity; ii) meso: number of clusters; iii) and micro: betweenness (Scott & Carrington, 2014). Density is the proportion of links in a network relative to the total number of links possible. The mean degree is the average number of links per node in the network. Modularity express a networks' strength of cluster division. Increasing values indicate the existence of a community-like structure. Clustering analysis identifies highly interconnected nodes to uncover known communities. Betweenness unveils a node's capacity in mediating the flow of information in a network. Increasing values indicate a higher betweenness. Bibliometrix (Aria & Cuccurullo, 2017) and Gephi (Bastian et al., 2009) were used for networks' layout and scores' computation.

**Results and discussion**

Figure 1 summarizes the macro, meso, and micro scores for each network. The period 2008-2012 stands out as the network with the highest increase in the number of nodes (198%), mean degree (55%), and clusters (39%). Density and modularity, however, diminished by 50% and 2%, respectively. The networks' density and modularity had decreased throughout 1998-2017 despite the mean degree's uninterrupted growth. The number of clusters peaked in 2008-2012 (39) and reached the lowest point in 2013-2017 (22).

Ronda and Guerras (2012) found that density and clusterability, a modularity proxy, have increased between 1962-2008 for the *strategic management* coword network. That seems plausible for an MBR sub-field. Using bibliographic coupling, Fanelli and Glänzel (2013) computed a mean degree and modularity for the field of *business and economics* of 27.3 and 0.5, respectively, which were similar to the mean degree of 2008-2012 (29.6) and the modularity of 2013-2017 (0.53). The caveat is that bibliographic coupling networks are different from co-word analysis, the latter similar to that of cocitations (Yan & Ding, 2012).

Figure 2 presents the four coword networks. Nodes' size is proportional to their betweenness. Nodes with labels are: i) those among the top-10 betweenness; ii) those that increased their position among the top-20 betweenness compared to the previous period (light-green colored); and iii) those that decreased their position among the top-20 betweenness compared to the previous period (red-colored). Clusters colored in burgundy and pine-green are the first and second-largest ones, respectively. I tagged clusters manually following a discernible thread among highly connected keywords. The two most crowded clusters for each period were: i) 1998-2002: *mathematical models, computer simulation, and algorithms* (18% nodes), and *science, technology and innovation* (STi) (10%); ii) 2003-2007: *LAC issues* (15%), and *strategic management for sustainable development* (15%); iii) 2008-2012: *STi in LAC* (10%), and *mathematical models, computer simulation, and algorithms* (9%); and iv) 2013-2017: *strategic management for sustainable development* (22%); and *logistics* (14%).

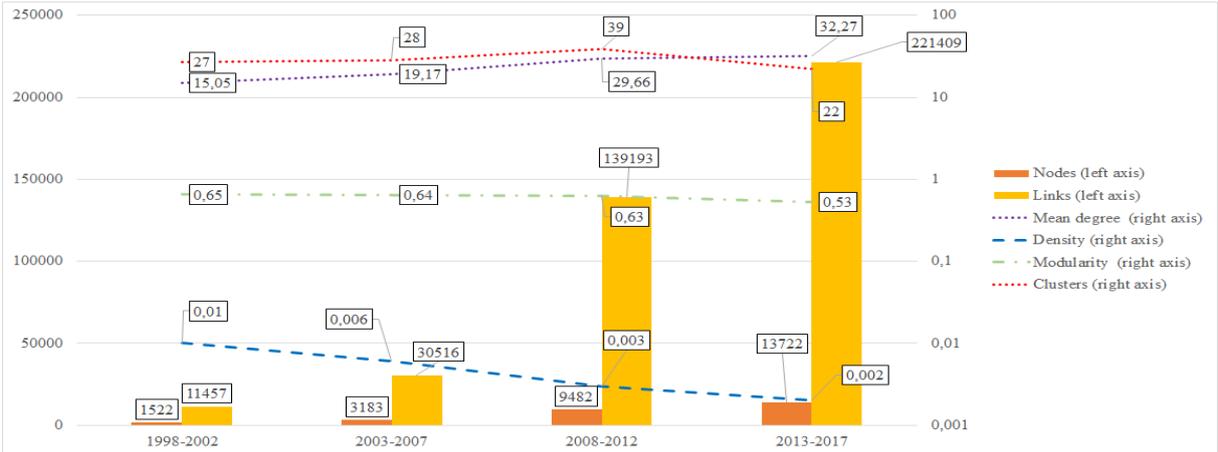

**Figure 1 Nodes, links, density, modularity, and clusters of coword networks 1998-2017. Source: elaborated by the author based on Scopus (2020).**

**1998-2002**          **2003-2007**

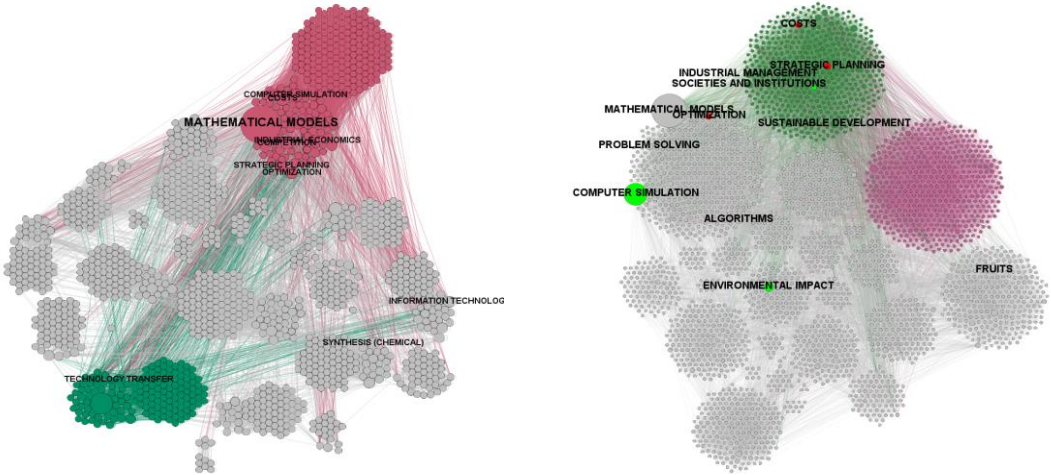

2008-2012                                              2013-2017

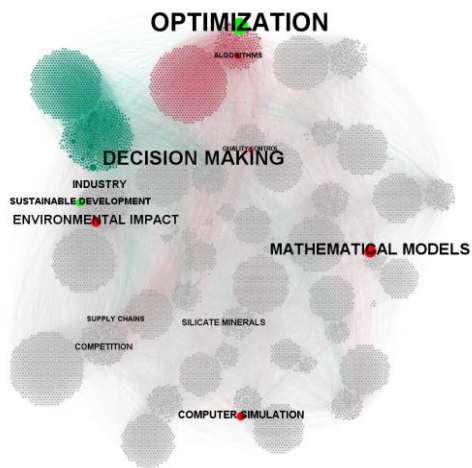 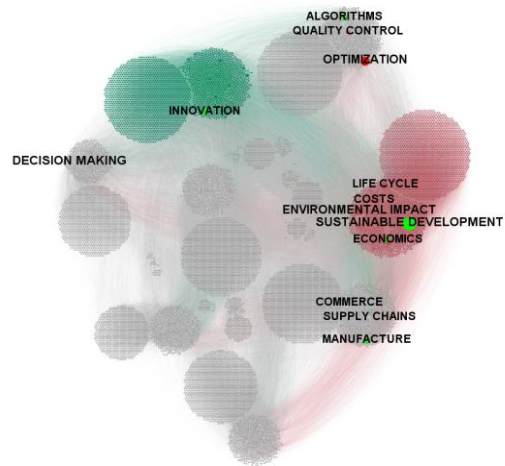

**Figure 2 Coword networks 1998-2017. Layout: circle park (hierarchy 1: modularity class; hierarchy 2: betweenness). Source: elaborated by the author based on Scopus (2020) and computed with bibliometrix (2017) for R and Gephi (2009).**

Table 2 (Appendix) presents the top-20 keywords by period according to their betweenness. Topics comprised of *mathematical models, computer simulation*, and *algorithms* cluster show consistent appearance among the four periods. *Sustainable development* and *environmental impact* related issues also showed consistent appearances. In contrast, MBR core topics, such as *strategic planning*, *quality control*, or *marketing*, showed either intermittent appearances or a downward trend.

Crowded clusters mixed between each other (e.g., STi *plus* LAC issues ⇒ STI in LAC) and specific keywords persisted (e.g., *mathematical models, computer simulation, and algorithms*; or *strategic management for sustainable development*). Contrasting those findings with the persistence and higher betweenness of the *mathematical models,* it unveils the broader applications of such methods for MBR and other disciplines (e.g., from supply chain management to e-commerce and genetics), even more than influential methods in MBR such as *case study* (actually, copied from medical sciences) (Eisenhardt, 1989). Furthermore, global solution-oriented agreements such as the Sustainable Development Goals produced inflections in the research output and permeated MBR in LAC and other developing regions (Cortés-Sánchez et al., 2020).

**Conclusion**

This study presented longitudinal evidence on the increasing dissension path of MBR in LAC. Also, it presented the mixture and persistence of crowded clusters and intellectual bridges. Such bridges were not from MBR but different disciplines moving toward consensus (e.g., mathematics or physics). Further research could implement non-redundant SM techniques such as social (i.e., coauthorships) or information (i.e., bibliographic coupling) based networks, discuss the advantage and obstacles of open vs. closed network structures, and source data from broader and inclusive bibliographic databases (e.g., Google Scholar or Dimensions).

**Appendix**

**Table 2 Top-20 keywords based on betweenness score 1998-2017**

| 1998-2002 | | 2003-2007 | | 2008-2012 | | 2013-2017 | |
|---|---|---|---|---|---|---|---|
| Keyword | Betweenness | Keyword | Betweenness | Keyword | Betweenness | Keyword | Betweenness |
| mathematical models | 2,62e+11 | mathematical models = | 1,10e+12 | optimization ↑ | 2,62e+12 | sustainable development ↑ | 6,52e+12 |
| costs | 5,61e+10 | computer simulation ↑ | 6,95e+11 | *decision making* | 2,15e+12 | decision making = | 6,02e+12 |
| computer simulation | 5,27e+10 | environmental impact ↑ | 2,28e+11 | mathematical models ↓ | 1,83e+12 | optimization ↓ | 4,12e+12 |
| strategic planning | 5,14e+10 | *industrial management* | 1,62e+11 | environmental impact ↓ | 1,48e+12 | environmental impact = | 3,65e+12 |
| argentina | 5,03e+10 | *problem solving* | 1,59e+11 | *industry* | 1,25e+12 | manufacture ↑ | 2,69e+12 |
| synthesis (chemical) | 4,35e+10 | *north america* | 1,56e+11 | computer simulation ↓ | 1,20e+12 | *life cycle* | 2,60e+12 |
| optimization | 4,07e+10 | *algorithms* | 1,55e+11 | sustainable development ↑ | 1,10e+12 | innovation ↑ | 2,37e+12 |
| technology transfer | 3,81e+10 | strategic planning ↓ | 1,40e+11 | *colombia* | 9,88e+11 | *costs* | 2,35e+12 |
| information technology | 3,59e+10 | *sustainable development* | 1,40e+11 | *competition* | 9,75e+11 | *supply chains* | 2,33e+12 |
| industrial economics | 3,29e+10 | *quality control* | 1,35e+11 | *silicate minerals* | 9,54e+11 | *commerce* | 2,28e+12 |
| environmental impact | 3,25e+10 | *data acquisition* | 1,33e+11 | *modeling* | 8,45e+11 | algorithms ↑ | 2,21e+12 |
| water | 3,19e+10 | *fruits* | 1,31e+11 | algorithms ↓ | 8,43e+11 | economics ↑ | 2,20e+12 |
| structural analysis | 3,17e+10 | marketing ↑ | 1,30e+11 | *production engineering* | 8,06e+11 | *environmental management* | 1,57e+12 |
| thermal effects | 3,06e+10 | *societies and institutions* | 1,24e+11 | *concentration (process)* | 7,29e+11 | *carbon dioxide* | 1,50e+12 |
| performance | 2,96e+10 | optimization ↓ | 1,13e+11 | *economics* | 7,20e+11 | *integer programming* | 1,32e+12 |
| marketing | 2,71e+10 | *food processing* | 1,09e+11 | *manufacture* | 7,03e+11 | *education* | 1,28e+12 |
| management | 2,58e+10 | costs ↓ | 1,06e+11 | quality control ↓ | 6,96e+11 | *regression analysis* | 1,25e+12 |
| raw materials | 2,56e+10 | *scheduling* | 9,66e+10 | *innovation* | 6,82e+11 | *investments* | 1,23e+12 |
| internet | 2,52e+10 | *systems analysis* | 9,13e+10 | *simulation* | 6,64e+11 | quality control ↓ | 1,21e+12 |
| venezuela | 2,34e+10 | *public policy* | 9,00e+10 | *developing countries* | 6,39e+11 | *fruits* | 1,19e+12 |

Note: new keywords compared to the former period are **bold** and *italic*. Symbols tell if the keyword increased (↑), diminished (↓), or maintained (=) its rank compared to the previous period. Source: elaborated by the author based on Scopus (2020) and computed with bibliometrix (2017) for R and Gephi (2009)